\documentclass[aps,prl,twocolumn,groupedaddress,amssymb,amsmath,nofootinbib]{revtex4}

\usepackage{color}
\usepackage{graphicx}
\usepackage{epstopdf}
\usepackage{hyperref} 

\begin{document}

\preprint{}

\title{
Comment on ``Supersoft Top Squarks"
}

\author{Radovan Dermisek}

\affiliation{Physics Department, Indiana University, Bloomington, IN 47405, USA}

\email[]{dermisek@indiana.edu}


\date{December 11, 2021}

\begin{abstract}

It is shown that the mechanism to soften contributions of top squarks to the Higgs mass, 
 discussed in ``Supersoft Top Squarks", Phys. Rev. Lett. \textbf{125}, no.15, 151801 (2020), 
is very similar, with identical result in the leading order, to the one discussed previously by the author in
``Loop Suppressed Electroweak Symmetry Breaking", Phys. Rev. D \textbf{95}, no.1, 015002 (2017). 
More general forms of Yukawa and soft mass squared matrices providing the same effect are also discussed.

\end{abstract}


\maketitle






Cohen, Craig, Koren, McCullough and Tooby-Smith~\cite{Cohen:2020ohi} have shown that the leading  contribution from top squark masses to the Higgs mass 
 in the renormalization group (RG) evolution from an arbitrary mediation scale $\Lambda_{UV}$ can be eliminated. Contributions to the Higgs mass are reduced to threshold corrections at the scale of new particles introduced in the model and other effects in the RG evolution resulting from gauge couplings. This possibility, with identical result in the leading order, was discussed previously in a model with the same particle content but with different structures of Yukawa and soft mass squared matrices~\cite{Dermisek:2016tzw}.  Furthermore, both models are examples of a whole class of models providing the same effect. We will identify  required structures of Yukawa and soft mass squared matrices.

The model in Ref.~\cite{Cohen:2020ohi} is the minimal supersymmetric model  extended by a copy of quarks $Q'$ and $U^{\prime c}$ with the same quantum numbers as the third generation standard model quarks $Q$ and $U^c$, and corresponding vectorlike partners $\bar Q'$ and $\bar U^{\prime c}$. The superpotential is given by 
\begin{equation}
W = \lambda_t H_u QU^c + \lambda_t H_u Q' U'^{c} + M(Q'\bar Q' + U^{\prime c}\bar U^{\prime c}),
\end{equation}
where the two Yukawa couplings are related by exchange symmetry. The soft masses of corresponding scalars are given by 
\begin{equation}
V_{Soft}  = \tilde m^2 ( |{\tilde Q}|^2 +  |\tilde  U^c|^2 -  |\tilde  Q'|^2 - |\tilde  U^{\prime c}|^2 ).
\end{equation}
The contributions of unprimed and primed fields to the Higgs soft mass squared in the RG evolution cancel as a result of identical Yukawa couplings and opposite sign soft masses squared. This cancellation is exact if gauge couplings and other Yukawa couplings are ignored. Negative soft masses squared of $\tilde Q'$ and $\tilde U^{\prime c}$ are not problematic because the supersymmetric mass $M$ can be chosen such that the resulting scalar masses squared are positive.

The model in Ref.~\cite{Dermisek:2016tzw} has the same field content, but only unprimed fields have sizable Yukawa coupling, and only primed fields have non-zero soft masses squared. This setup again leads to zero contribution to the Higgs mass  in the RG evolution since the fields with non-zero soft masses do not couple to the Higgs boson at all. The supersymmetric masses can be chosen such  that the resulting spectrum is again phenomenologically viable. 

Contributions to the Higgs mass  in both models are reduced to threshold effects at the scale $M$ (replacing $\Lambda_{UV}$), and other effects in the RG evolution resulting from gauge couplings: gaugino  and 2-loop scalar contributions, that could still be sizable depending on other assumptions~\cite{Cohen:2020ohi, Dermisek:2016tzw}.

In order to identify more general forms of Yukawa couplings and soft masses leading to the same effect, it is instructive to write the relevant Yukawa terms in the 1-loop RG equation for $m^2_{H_u}$ in a matrix notation~\cite{Martin:1993}:
\begin{equation}
\beta_{m^2_{H_u}}  \supset \; 6\, {\rm Tr} \left[{\bf m_Q^2 Y_u^\dagger Y_u} + {\bf Y_u^\dagger  m_U^2 Y_u} \right],
\end{equation}
where $\bf Y_u$ is the matrix of Yukawa couplings to $H_u$ in the basis $(Q,Q')$ and $(U^{c},U'^{c})$, and $\bf  m_{Q,U}^2$ are matrices of corresponding soft squark masses squared. 

The scenario in Ref~\cite{Cohen:2020ohi} corresponds to
\begin{equation}
{\bf Y_u}  \; = \; \lambda_t \,\left(\begin{array}{cc} 
                                             1 & 0 \\
                                             0 & 1  
\end{array}   \right),\quad 
{\bf m_{Q,U}^2}  \; = \; \tilde m^2 \left(\begin{array}{cc} 
                                             1 & 0 \\
                                             0 & -1  
\end{array}   \right),
\label{eq:S1}
\end{equation}
%
while the scenario discussed in Ref~\cite{Dermisek:2016tzw} has
\begin{equation}
{\bf Y_u}  \; = \; \lambda_t \,\left(\begin{array}{cc} 
                                             1 & 0 \\
                                             0 & 0  
\end{array}   \right),\quad 
{\bf m_{Q,U}^2}  \; = \; \tilde m^2 \left(\begin{array}{cc} 
                                             0 & 0 \\
                                             0 & 1  
\end{array}   \right). 
\label{eq:S2}
\end{equation}
Clearly, since both structures of $\bf Y_u$ are RG invariant (including also gauge interactions) and structures of ${\bf m_{Q,U}^2} $ are maintained in the RG evolution by  corresponding $\bf Y_u$, both scenarios lead to zero contribution to $m^2_{H_u}$.


The scenario in Eq.~(\ref{eq:S1}) could be further generalized to:
\begin{equation}
{\bf Y_u}  \; = \; \lambda_t \, \left(\begin{array}{cc} 
                                             1 & a \\
                                             a & 1  
\end{array}   \right),\quad 
{\bf m_{Q,U}^2}  \; = \; \tilde m^2_{Q,U} \left(\begin{array}{cc} 
                                             1 & a_{Q,U} \\
                                             -a_{Q,U} & -1  
\end{array}   \right),
\label{eq:S3}
\end{equation}
where $a$, $a_{Q}$, and $a_{U}$ are not related.
These forms of $\bf Y_u$ and ${\bf m_{Q,U}^2}$ do not change in the RG evolution. Their products remain traceless and thus  the correction to $m^2_{H_u}$ from the RG evolution remains zero. Similarly, the scenario in Eq.~(\ref{eq:S2}) could be further generalized to include arbitrary off-diagonal elements in ${\bf m_{Q,U}^2}$. The generalized scenario in Eq.~(\ref{eq:S3}) reduces to the one in Eq.~(\ref{eq:S1}) for $a=0$ (up to off-diagonal entries in ${\bf m_{Q,U}^2}$ that do not contribute to the Higgs mass), and for $a=1$ the $\bf Y_u$ is equivalent (up to a change of basis) to $\bf Y_u$ in Eq.~(\ref{eq:S2}) with soft masses having one of the generalized form.


\vspace{0.5cm}
\noindent
{\bf Acknowledgments:} R.D. thanks M.~McCullough for correspondence. This work was supported in part by the U.S. Department of Energy under grant number {DE}-SC0010120.



\end{document}